\documentclass[published]{JHEP3} 
\JHEP{04(2004)038}
\received{April 1, 2004}
\accepted{April 18, 2004}

\skip\footins = 1\bigskipamount plus 2pt minus 4pt

\usepackage{epsfig}

\newcommand\U{\mathop{\rm {}U}\nolimits} 

\title{Regular S-brane backgrounds}

\author{Gianmassimo Tasinato\\
Physikalisches Institut der Universit\"at Bonn\\
Nussallee 12, 53115 Bonn, Germany\\
E-mail: \email{tasinato@th.physik.uni-bonn.de}}

\author{Ivonne Zavala\\
Department of Physics, University of Colorado\\
Boulder, CO, 80309, U.S.A.\\
E-mail: \email{zavala@pizero.colorado.edu}}

\author{Clifford P. Burgess\\
Physics Department, McGill University\\  
3600 University Street, Montr\'eal, Qu\'ebec, Canada, H3A 2T8\\
E-mail: \email{cliff@physics.mcgill.ca}}

\author{Fernando Quevedo\\
Centre for Mathematical Sciences, DAMTP, University of Cambridge\\
Cambridge CB3 0WA U.K.\\
E-mail: \email{f.quevedo@damtp.cam.ac.uk}}

\abstract{We construct time-dependent S-brane solutions to the
  supergravity field equations in various dimensions which (unlike
  most such geometries) do not contain curvature singularities.  The
  configurations we consider are less symmetric than are earlier
  solutions, with our simplest solution being obtained by a simple
  analytical continuation of the Kerr geometry. We discuss in detail
  the global structure and properties of this background. We then
  generalize it to higher dimensions and to include more complicated
  field configurations --- like non vanishing scalars and antisymmetric
  tensor gauge potentials --- by the usual artifice of applying
  duality symmetries.}

\begin{document}

\section{Introduction and summary}

\subsection{Introduction}

Over the past years, considerable effort has been devoted to
understanding the issue of time dependence in string theory. A
practical place to begin the search for time-dependent string
configurations is to find solutions to the field equations which
describe the low-energy limit of string theory. The search for such
solutions has received special attention, and has led to the study of
the so called space-like branes or S-brane
solutions~\cite{strominger}.\footnote{See~\cite{oldwork} for early
  work on this subject.} Besides providing time-dependent backgrounds
which have a string pedigree, the S-branes --- embedded in
asymptotically flat, time-dependent backgrounds --- are believed to be
relevant for describing the dynamics of tachyon fields~\cite{tachyon}.

Time-dependent configurations have an obvious appeal as potentially
having cosmological applications~\cite{cornalba,wohlfart}.  In
particular, because the induced metric which observers embedded within
these space-times can experience do not satisfy the usual Friedmann
equation, they can undergo accelerated \pagebreak[3] expansion, or
bouncing~\cite{rabadan}, without necessarily requiring the problematic
kinds of matter which would normally be required.  Since these
spacetimes are time-dependent, they provide laboratories for studying
particle production, such as has been considered in~\cite{bmqtz}.

A large class of time dependent, asymptotically-flat S-brane solutions
have been discussed in the literature~\cite{othersbranes,bqrtz}. A
common feature which these space-times typically share is that they
are plagued by singularities. These can arise as null singularities,
or naked time-like singularities inside internal static regions. This
observation has led to the development of a theorem, which states that
such singularities are inevitable for a fairly broad class of
metrics~\cite{buchel} (see also~\cite{leblond}).

\looseness=-1 Although it may ultimately be possible to resolve these
singularities within string theory, in this paper we instead ask
whether singularity-free configurations can be found purely within the
low-energy limit. We show that they can, by starting with a more
general ansatz for the metric; a similar observation has been recently
made in~\cite{stromreg,wang}.  The resulting exact solutions are
typically less symmetric than the solutions already in the literature,
but reduce to them in the limit where the asymmetries in the solution
go to zero.

\subsection{Summary}

Our presentation comes in two parts. In the first one,
section~\ref{section2}, we examine the properties of the simplest
non-trivial, non-singular, time-dependent solution. In Part II this
solution is generalized in several directions.

\paragraph{A simple solution.} In the first part, we obtain the 
simplest solution by performing a suitable analytical continuation to
the Kerr solution, in the same way that the well-known
four-dimensional S0-brane solution,
\begin{equation}\label{ssbs}
    d s^2  = - \frac{dt^{2}}{(1-{2 \,m}/{t})}
    +\left( 1-\frac{2 \,m}{t} \right)  d r^{2} +t^{2}
    d {\mathcal H}_2^{2}\,,
\end{equation}
can obtained by the same kind of continuation from the static,
spherical Schwarzschild black hole~\cite{bqrtz}. Here $d {\mathcal
  H}_2^2$ is the metric of the transverse spatial sections, which have
a hyperbolic symmetry. This solution has an asymptotically-flat
time-dependent region which is separated by a Cauchy horizon from an
internal, static region containing a naked time-like singularity.

Performing a similar continuation of the Kerr rotating black hole
solution produces a time-dependent spacetime having less symmetry, but
which is completely regular and without curvature
singularities.\footnote{The observation that angular momentum can cure
  singularities in string theory was noticed in~\cite{maoz}.}  The
metric we would obtain is:
\begin{eqnarray}\label{kerrSbrintr}
    d s^{2} &=& -\frac{\Sigma}{\Delta}
    \,dt^{2} + \left( \frac{\Delta+a^{2} \sinh^{2}\theta}{\Sigma}
    \right)d r^{2}+\Sigma \,d\theta^{2}+ 
\nonumber \\
    &&+
    \left[(t^2+a^{2})^{2}+a^{2} \sinh^{2}\theta \Delta  \right]
    \frac{\sinh^{2}\theta}{\Sigma} \,d \phi^{2}
    -\frac{4 m\, a\, t}{\Sigma}\sinh^{2}\theta\, d \phi \, d r\,,
\end{eqnarray}
where
\begin{eqnarray*}
    \Sigma &=& t^{2}+a^{2} \cosh^{2}\theta\,, 
\\
    \Delta &=& t^{2}-2 m t +a^{2}=(t-m)^{2}+a^2-m^2 \,.
\end{eqnarray*}
The solution depends on the parameter $a$, an integration constant
which is the analog of the angular momentum of the Kerr solution, and
a parameter $m$, that in the black hole case corresponds to the
mass. The nature of the solution depends crucially on the value of $a$
and $m$.

As we will see, when $a=0$, we recover the S-brane solution of
eq.~(\ref{ssbs}), together with its singularity. However, as soon as
$a$ is nonzero, but small, the solution that results is free of
curvature singularities. It instead has a stationary region which
contains closed time-like curves (CTC's). This region is separated by
Cauchy horizons, from external time-dependent regions.

More interestingly, for large enough values for the parameter $a$ the
geometry is \emph{completely} regular and causally well-behaved. It is
defined for all values of the coordinates and it does not contain
horizons or singularities. In particular, we show that the geodesics
are well-behaved everywhere.
                       
\paragraph{Generalizations.} In the second part,
section~\ref{section3}, we generalize this simple geometry to more
complicated situations. We do so in several ways. One non trivial
extension is to $d$ dimensions higher than four, where the ($d-4$)
extra dimensions are characterized by a hyperbolic symmetry.  We also
find new solutions by T-dualising the simpler configuration we found
in the first part, along the lines of~\cite{cliffrob}.  Other
solutions are generated using similar solution-generating techniques
in higher dimensions, including two examples of configurations with
non-trivial dilaton and antisymmetric form fields in five
dimensions. The first of these examples is obtained by applying
T-duality plus coordinate rotations~\cite{hhs} to a known
five-dimensional vacuum spacetime, leading to regular configurations
with non-trivial NS fields. The second example applies a Hassan-Sen
transformation~\cite{hassansen} to a known solution, leading to a
nontrivial time-dependent solution for low-energy heterotic string
theory with nonzero one- and two-form fields. For each configuration,
we identify the regions of parameter space for which our
time-dependent solutions are completely regular.

\section{Kerr S-branes}\label{section2}

In this section we construct and discuss in detail the properties of the
simplest regular S-brane solution. This is obtained as an analytical
continuation of the Kerr black hole in vacuum four-dimensional
Einstein gravity. We begin with a review of the Kerr black hole,
followed by the construction of its S-brane generalization --- the
S-Kerr solution --- and a study of its properties.

\subsection{The Kerr geometry}

The four dimensional Kerr black hole is the axially-symmetric solution
to the vacuum Einstein equations which describes the spacetime outside
of a rotating black hole. In Boyer-Lindquist coordinates, the solution
is given by
\begin{equation}\label{statkerr}
    d s^{2}=- d t^{2} + \Sigma \left(\frac{d r^{2}}{\Delta}
    + d \theta^{2} \right)+
    (r^2+a^2) \sin^{2}\theta d \phi^{2} +\frac{2 m r}{\Sigma}
    \left(a \sin^{2}\theta d \phi -d t \right)^{2},
\end{equation}
where
\begin{eqnarray}
    \Sigma &=& r^{2}+a^{2} \cos^{2}\theta 
\nonumber  \\
    \Delta &=& r^{2}-2 m r +a^{2}=(r-m)^{2}+a^2-m^2 \,.
\end{eqnarray}
In these expressions, $m$ represents the mass of the black hole and
$a\equiv J/m$ parameterizes the angular momentum per unit mass. This
metric has the following properties on which we shall draw in what
follows (see, $e.g.$, ref.~\cite{hawking} for details):
\begin{itemize}
\item It is asymptotically flat, as an examination of the Riemann
  tensor shows. The variable $\phi$ parameterizes the direction of the
  axial symmetry, and can be compactified to lie in the interval
  $[0,2\pi]$ without introducing spurious conical singularities
  anywhere.
\item It has two isometries, corresponding to the shifts in the
  coordinates $t$ and $\phi$ in Boyer-Lindquist coordinates. These
  correspond physically to time-translation invariance and axial
  symmetry.  There is indeed a mixed term $g_{t\phi}$ indicating that
  the solution rotates.
\item It has a curvature singularity where the invariant $R^{\mu \nu
  \lambda \sigma} R_{\mu \nu \lambda \sigma}$ diverges. This occurs
  when $\Sigma=0$, or equivalently when $r=0$ and $\theta=\pi/2$
  simultaneously. This is the standard ring singularity of the Kerr
  geometry.
\item It has coordinate singularities when $\Delta=0$, which turn out
  to be horizons of the Kerr geometry. When $a^{2}< m^{2}$ the metric
  has two horizons.\footnote{Cosmological aspects of the time
    dependent region inside the two horizons has been discussed
    in~\cite{octavio}.}  The internal one hides a ring singularity
  located at $\Sigma=0$, that is for $r=0$ and $\theta=\pi/2$.  There
  is also an ergoregion which is determined by the zeroes of $g_{tt}$,
  where the infinite red shift surfaces are located. The outer
  boundary of the ergosphere is located outside the outer horizon.  On
  the other hand, when $a^{2} > m^{2}$, the metric does not have
  horizons and the ring singularity is naked.
\item The Kerr geometry has closed time-like curves, for small
  negative values of $r$. When $a^2<m^2$, these are safely limited to
  a stationary region inside the event horizon.
\end{itemize}
These properties change dramatically for the Kerr S-brane, as we now
show.

\subsection{The Kerr S-brane}\label{simplerot}

In order to construct the simplest Kerr S-brane solution we follow the
steps taken to obtain the more symmetric S-brane solutions from the
simpler Schwarzschild geometry. This is done via a suitable analytical
continuation of the Schwarzschild black hole introduced
in~\cite{bmqtz,bqrtz}, leading to a spacetime with time-like naked
singularities. In the same spirit we perform the following analytic
\emph{i}-rotation~\cite{bqrtz} in the Kerr geometry:
\begin{equation}
    t \to i r\,,
\qquad 
r \to it\,, 
\qquad 
\theta \to i\theta\,,
\qquad 
a \to ia\,, 
\qquad 
m \to im \,.
\end{equation}
This leads to the following time-dependent, axially symmetric solution
to the four dimensional Einstein equations in vacuum:
\begin{eqnarray}\label{kerrSbr}
    d s^{2} &=&  -\frac{\Sigma}{\Delta} \,dt^{2}+ \left( \frac{\Delta+a^{2}
    \sinh^{2}\theta}{\Sigma}
    \right)d r^{2}+\Sigma \,d\theta^{2}+ 
\nonumber \\
    &&+
    \left[(t^2+a^{2})^{2}+a^{2} \sinh^{2}\theta \Delta  \right]
    \frac{\sinh^{2}\theta}{\Sigma} \,d \phi^{2}
    -\frac{4 m\, a\, t}{\Sigma}\sinh^{2}\theta\, d \phi \, d r\,,
\end{eqnarray}
where now
\begin{eqnarray}
    \Sigma &=& t^{2}+a^{2} \cosh^{2}\theta\,, 
\nonumber \\
    \Delta &=& t^{2}-2 m t +a^{2}=(t-m)^{2}+a^2-m^2 \,.
\end{eqnarray}

The solution~(\ref{kerrSbr}) inherits two isometries from the Kerr
solution, corresponding to shifts in the coordinates $r$ and $\phi$.
The mixed term $g_{t\phi}$, indicating the rotational structure of the
Kerr black hole, has now been replaced by a mixed term $g_{r\phi}$,
indicating a ``rotation'' in space, or, better, a helical or wrung
structure.  The solution is also invariant under the exchange $a \to
-a$ and $m \to -m$, a transformation whose physical interpretation we
discuss in the next section.  Notice too that we recover the
well-known Schwarzschild \emph{i}-rotated solution or S0-brane
(complete with its time-like naked singularities at the origin) in the
limit $a \to 0$.

The key property of this new solution follows from the observation
that, unlike for the Kerr geometry, the quantity $\Sigma$ cannot be
zero if $a \ne 0$. As a consequence there is no analog of the Kerr
ring singularity for this new solution. For instance, the curvature
invariant $R_{\mu \nu \lambda \sigma} R^{\mu \nu \lambda \sigma}$ is
given by the following expression\footnote{We thank 
John Wang for pointing us out some typos in this formula in a
previous version of the paper.}:
\begin{equation}\label{riccinv}
    R_{\mu \nu \lambda \sigma} R^{\mu \nu \lambda \sigma} =
\frac{48 m^2}{\Sigma^6}
      [t^2-a^2\cosh^2\theta][t^4-14\,a^2\, t^2\cosh^2\theta +
 a^4\cosh^4\theta]
\end{equation}
{}from which it is clear that this invariant is well defined
everywhere so long as $a \neq 0$.\footnote{In~\cite{poritz}, a similar
  procedure of analytic continuation has been used to obtain regular
  time-dependent solutions in the presence of a cosmological
  constant.}

\subsection{Global structure}	

We now examine the global structure of the Kerr S-brane in more
detail.

\paragraph{Asymptopia.} As might be expected, the Kerr 
S-brane inherits an asymptotically flat region from the Kerr black
hole. This property is less evident in the Boyer-Lindquist form of the
metric used above due to the presence of the diverging $\sinh$ and
$\cosh$ functions. To display this region more explicitly it is
convenient instead to pass to Kerr-Schild coordinates, in terms of
which the solution may be rewritten as
\begin{equation} 
\label{kerrschild}
d s^{2}=d \bar{r}^{2}+ d x^{2}+ d y^{2}- d z^{2} -
\frac{2m t^{3}}{t^{4}+a^{2} z^{2}} \left( \frac{t(x dx +y d y)
-a (x d y- y d x)}{t^{2}+a^{2}}+\frac{z d z}{t}+d \bar{r}
\right)^{2} ,
\end{equation}
where
\begin{eqnarray}
x+iy &=& (t+ia) \sinh\theta \exp
\left[ i \int (d \phi+ a \Delta^{-1} d t)\right] 
\nonumber \\
z &=& t \cosh\theta \,,
\qquad 
\bar{r} =\int(d r+(t^{2}+a^{2})\Delta^{-1} d t) -t \,.
\end{eqnarray}
Here $t$ is to be regarded as being a function of the other
coordinates, as determined implicitly from the condition
\begin{equation}
    t^{4}-(z^{2}-x^{2}-y^{2}-a^{2})t^{2}-a^{2}z^{2}=0\,.
\end{equation}
From these expressions it is easy to see that for large values
of the coordinates the geometry  approaches the flat space, since
in this limit the second line of eq.~(\ref{kerrschild}) vanishes.

\paragraph{Location of horizons.} Although we have seen 
that the metric does not have curvature singularities, in its
form~(\ref{kerrSbr}) it does have coordinate singularities for those
points where the function $\Delta$ vanishes, which happens when
\begin{equation} \label{horizcond}
    t = t^{h}_{\pm} \equiv m \pm \sqrt{m^{2}-a^{2}}\,.
\end{equation}
These represent horizons of the geometry.  From eq.~(\ref{horizcond})
we see that the global properties of the geometry depend crucially on
the relative size of the parameters $a$ and $m$, and so we now discuss
the various cases separately.

\subsubsection{The case $a^2 < m^2$: closed timelike curves}

\FIGURE[b]{\epsfig{file=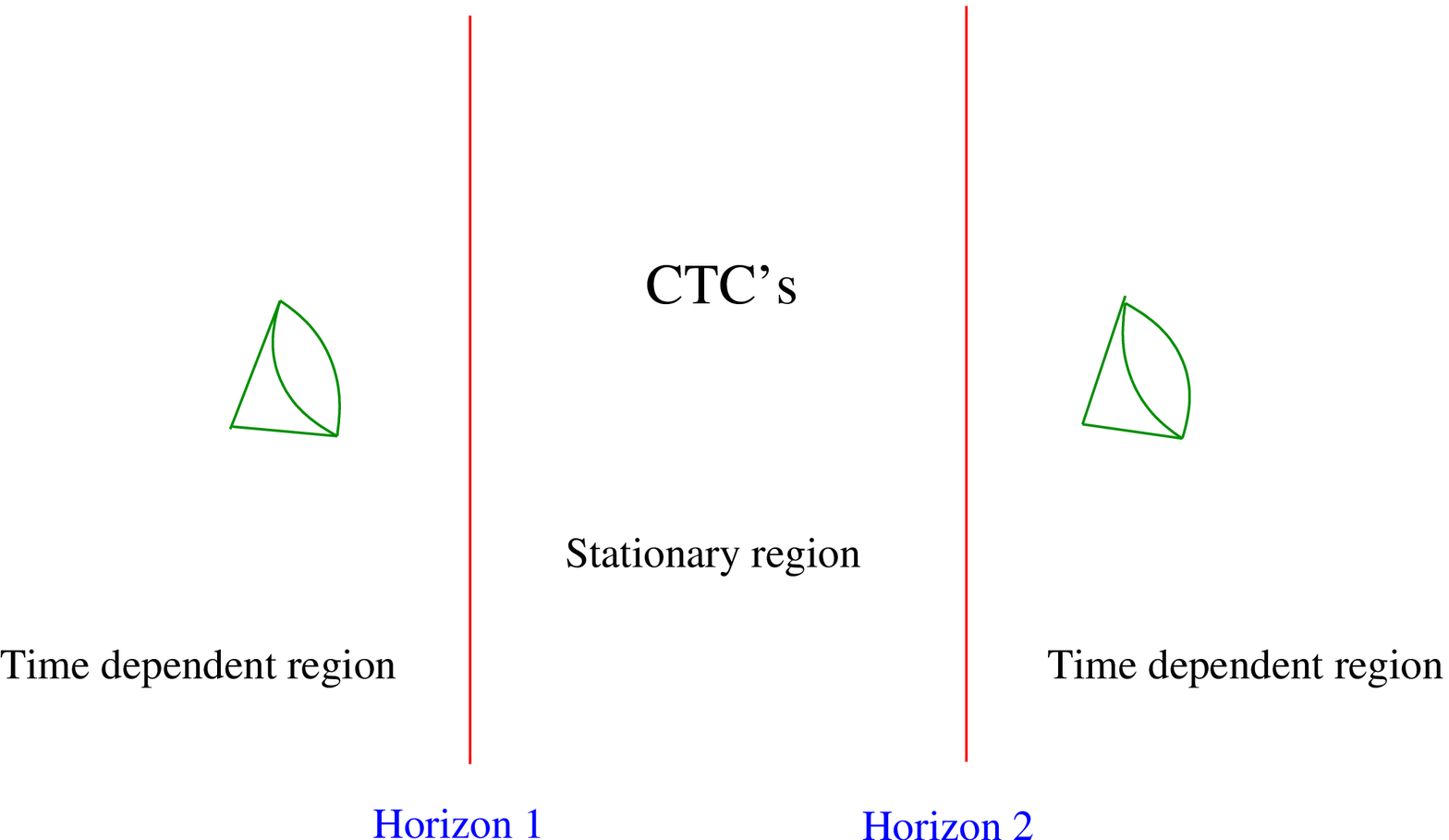, width=.7\textwidth}%
\caption{Pictorial representation of the Kerr S-brane
  solution~(\ref{kerrSbr}), when $a^2<m^2$. In the external, time
  dependent regions, we draw examples of future light cones. Notice
  that the internal, stationary region contains CTC's, making it
  meaningless to represent future light cones. \label{artpict}}}

In this case the configuration has two horizons, which separate two
asymptotically-flat and time-dependent regions from a stationary
region which contains closed timelike curves. This situation is
illustrated in figure~\ref{artpict}. If $ a^2 = m^2$ then the two
horizons degenerate into one. Notice that the two external
time-dependent regions get interchanged with one another under the
transformation $a \to -a$ and $m \to -m$, which leaves the metric
invariant.

The absence of a time-like curvature singularity for $0<a^2<m^2$, with
its replacement by a second horizon, allows a new interpretation of
the time-like singularity which appears in the limiting case $a =
0$. It suggests that the time-like singularity, found in this limit
might be thought of as a second S-brane, (that is, an
asymptotically-flat and time-dependent region bounded by a horizon),
but in the limit where its horizon collapses to zero size and becomes
singular. If $a$ is small but nonzero, then the singular horizon
becomes regular and the geometry describes two S-branes separated by
the internal stationary region.  In this case, there is also an
equivalent of the Kerr's ergoregion, limited by the boundaries located
at the points where $\Delta + a^2\,\sinh\theta =0 $. The outer of
these surfaces is located inside the first horizon, in contrast with
the Kerr black hole case.

Although there are no curvature singularities, the stationary interior
region of the geometry for $a^2 < m^2$ contains closed timelike
curves, and this means that it is not causally well defined. The
existence of these curves arises because in the stationary region it
is the coordinate $\phi$ which becomes time-like,\footnote{A simple
  closed timelike curve is then obtained by moving along trajectories
  on which only $\phi$ varies, given that $\phi$ is a periodic
  variable with period $2 \pi$.  However, notice that the presence of
  CTC's does \emph{not} rely on the periodicity of $\phi$
  (see~\cite{carter} for more details).}  since the coefficient of $d
\phi^{2}$ in~(\ref{kerrSbr}) becomes negative:
\begin{equation}
\label{condCTC}
\left[ (t^2+a^{2})^{2}+a^{2} \sinh^{2}(\theta) \Delta\right] \le 0\,.
\end{equation}
This condition can be satisfied whenever $\Delta < 0$ --- that is,
\emph{only} in the stationary part of the solution, inside the Cauchy
horizons.  Nevertheless, since there are no event horizons hiding
these causally-problematic regions from external observers, it is not
clear that the geometry in the case $a^2<m^2$ has a sensible physical
interpretation~\cite{hawkingcrono}.\footnote{As we will see in the
  second part of this letter, also adding other fields motivated by
  string theory, we find regular solutions with CTC's. Although it is
  not yet clear whether a chronology protection agency exists in
  string theory, there are some indications that this might be indeed
  the case (see e.g.~\cite{cronology}).}

\subsubsection{The Case $a^2 > m^2$: Nonsingular Geometries}

\FIGURE[t]{\epsfig{file=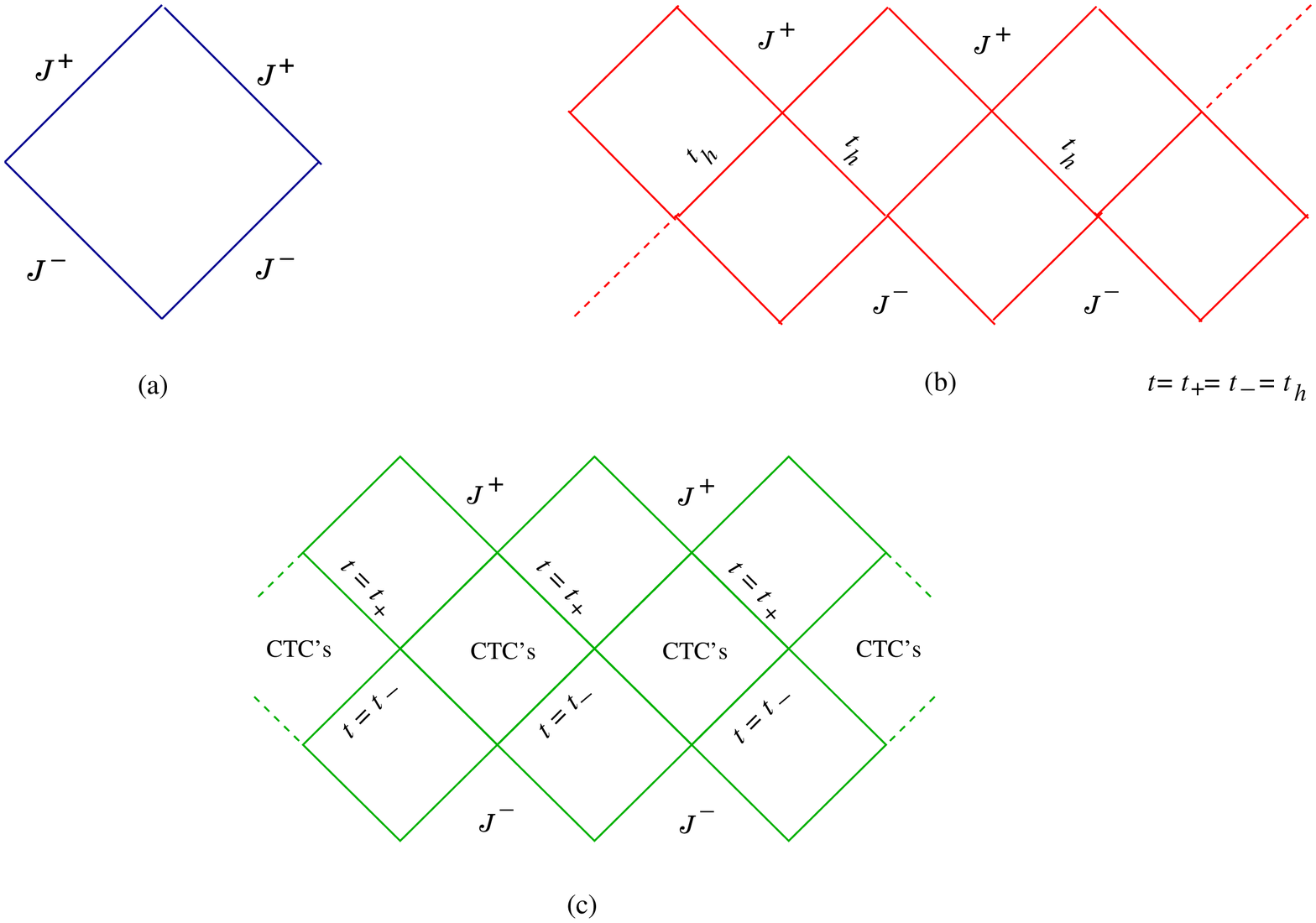, width=.9\textwidth}%
\caption{Penrose diagrams for the S-Kerr brane for $(a)$ $a^2>m^2$,
  $(b)$ $a^2=m^2$, $(c)$ $a^2<m^2$. \label{penroses}}}

When $a^2 > m^2$ neither $\Sigma$ nor $\Delta$ can vanish, and so the
geometry we obtain is well defined and time-dependent everywhere, and
asymptotically approaches flat space. So long as $m \ne 0$ the
geometry is curved, because its Riemann tensor does not
vanish.\footnote{When $m=0$ the geometry is flat, being Minkowski
  space-time in an unusual coordinate system (as is most apparent by
  choosing $m=0$ in formula~(\ref{kerrschild})).} The resulting
configuration is time dependent and does not contain horizons or
curvature or metric singularities. As we now show, the geodesics for
this geometry are perfectly well defined.  The Penrose diagrams along
the symmetry axis (that is, for $\theta=0$) for all the solutions are
presented in figure~\ref{penroses}.

\paragraph{Geodesic motion.} Let us now consider the geodesics 
of the time dependent geometry for $a^2 >m^2$. We restrict ourselves
here to the simplest case for which the coordinate $\theta$ does not
vary.

Due to the two isometries there are two conserved quantities which
provide two first-integrals of the geodesic equations. These can be
defined by
\begin{eqnarray}
-P &=& \left(\frac{2 m t}{\Sigma}-1\right)\dot{r}
+\frac{2 m t}{\Sigma} a \sinh^{2}{\theta} \dot{\phi}\,, 
\nonumber \\
L &=& - \frac{2 m t}{\Sigma} \,
 a \sinh^{2}{\theta} \,\dot{r}+\left[
(t^2+a^{2})^{2}+a^{2} \sinh^{2}\theta \Delta \right]
 \frac{\sinh^{2}\theta}{\Sigma}\, \dot{\phi}\,.
\end{eqnarray}
Where a dot means derivative with respect to the geodesic affine
parameter.  In terms of these the geodesic equations can be written as
follows:
\begin{eqnarray}
  \epsilon &=&  -\frac{\Sigma}{\Delta}
  \dot{t}^{2}+\frac{\Delta}{\Sigma} \left( \dot{r} +a \sinh^2\theta
  \dot{\phi}  \right)^{2} - \frac{\sinh^2{\theta}}{\Sigma}\left[
 (t^{2}+a^{2}) \dot{\phi} -a \dot{r} \right]^{2}  ,
   \\ 
0 &=& -\frac{a^{2}}{\Delta} \dot{t}^{2} -\frac{a^{2} \Delta}{\Sigma^{2}} 
\left( \dot{r} +a \sinh^2{\theta} \,\dot{\phi}  \right)^{2}
    +\frac{2 a \Delta \dot{\phi}}{\Sigma} \left( \dot{r} -a \sinh^2{\theta}\,
    \dot{\phi} \right) +
\nonumber \\
 && + \frac{(t^{2}+a^{2})}{\Sigma^{2}}\left[ (t^{2}+a^{2}) \dot{\phi}
  -a \dot{r} \right]^{2} ,
\end{eqnarray}
where $\epsilon=0 \,(-1,+1)$ for null (time-like, space-like)
geodesics. Simple algebraic manipulations reveal that, starting from
the previous equations characterizing the geodesics, the invariants
$P$ and $L$ satisfy the following relation:
\begin{equation}\label{simplman}
    (L-a P \sinh^2{\theta}) (L+a P
    \sinh^2{\theta}) = - \epsilon \,a^{2} \sinh^4{\theta} \,.
\end{equation}
We now discuss the behavior of null and time-like geodesics in more
detail.

\paragraph{Null Geodesics.}
In this case, the relation~(\ref{simplman}) considerably simplifies
the equations of motion. Using the condition
\begin{equation}
    (L-a P \sinh{\theta}^{2})=0\,,
\end{equation}
the geodesic equations simplify to the following expressions for the
first derivatives along the geodesic affine parameter:
\begin{equation}
    \dot{t} = \pm P \,, 
\qquad
    \dot{r} = \frac{(t^2+a^2)}{\Delta} \,P  
\qquad
    \hbox{and} 
\qquad
    \dot{\phi} = \frac{a P}{\Delta}\,. 
\end{equation}
These first-order differential equations are well-defined for any
value of the coordinates and of the affine parameter. Because $\Delta$
cannot vanish, these equations can be integrated without introducing
singularities, leading to geodesics which are also everywhere
well-defined, making the space-time geodesically complete for null
geodesics.

\paragraph{Time-like geodesics.}
The equations for timelike geodesics are harder to handle. We
concentrate on the simplest situation, for which the constant
coordinate $\theta$ is chosen to vanish: $\theta=0$. We obtain in this
case
\begin{eqnarray}\label{tlgeo}
    \dot{t} &=& \pm\left( P^{2} + \frac{\Delta}{\Sigma} \right), 
\nonumber\\
    \dot{r} &=& \frac{\Sigma}{\Delta} \,P\,,
\nonumber\\ 
    \dot{\phi} &=& \frac{a P}{\Sigma} \left(\frac{\Sigma}{\Delta}-1 \right)
    \left[1\pm \sqrt{1+\frac{\Delta^{2} \Sigma}{P^{2}(\Delta-\Sigma)^{2}}
    \left[\frac{P^{2}}{\Delta} \left(1+\frac{1}{a}\right) +\frac{1}{\Sigma}
    -\frac{P^{2}\Sigma}{\Delta} \right] }\, \right] .\qquad
\end{eqnarray}
It is also true in this case that the geodesics which result from
these equations are well-defined for all values of the coordinates,
and the space time is geodesically complete for these time-like
geodesics.

\section{Generalizations}\label{section3}

In this section we generalize in various ways the time dependent Kerr
S-brane solution of the last section. We first extend the four
dimensional pure-gravity solution to a more general solution of the
combined Einstein-dilaton-antisymmetric tensor equations of 4D
supergravity, for which the dilaton and antisymmetric tensor fields do
not vanish. This new solution is obtained by performing a T-duality
transformation, exploiting the T-duality invariance of the
supergravity action.

Next we extend the solutions in the vacuum to higher dimensions: we
obtain these solutions by performing a complex substitution in
stationary solutions already known in the literature.

We then continue with two examples, which show how to generalize
higher-dimensional solutions by acting on them with symmetries.
Similar to the 4D case after T-duality, the new time-dependent
solutions obtained in this way contain nontrivial dilaton, and are
coupled to two-forms and/or one forms.

We obtain the first of these higher-dimensional examples by performing
a coordinate rotation, followed by a proper T-duality transformation,
leading to a 5-dimensional S-string configuration, involving a
background dilaton field and an antisymmetric three form field.  In
the second example, we start with an S-string in five dimensional
vacuum, and, by means of applying a Hassan-Sen transformation, we
extend it to a solution within a more general background, containing
in addition nonzero abelian gauge fields.  Their global properties
are richer than the four dimensional ones, and we discuss
them in some detail.

\subsection{Four dimensional T-dual configurations}\label{Tdual4}

Although our Kerr S-brane configuration so far has been considered as
a solution to the vacuum Einstein equations, it may also be regarded
as a solution to the combined Einstein-Dilaton-Antisymmetric tensor
field equations of supergravity models, for which the dilaton, $\Phi$,
is a constant and the antisymmetric tensor field, $B_{\mu\nu}$,
vanishes. The action for this model, in the string frame, is given
explicitly by
\begin{equation}\label{het4dim}
    S= -\int dx^4 \sqrt{-g_4} \,e^{-\Phi}\left[\, R
    + (\nabla\Phi)^2 - \frac{1}{12}\,H^2_\emph{3}\right] ,
\end{equation}
where the 3-form $H_\emph{3}$ is related to the 2-form potential,
$B_\emph{2}$, by $H_\emph{3} = \hbox{d} B_\emph{2} + \dots$, where the
dots denote possible Chern-Simons terms which play no role in what
follows.

Given any solution to the field equations of this action which enjoy
at least one continuous symmetry, there is a well-known prescription
for generating new solutions. The new solution is obtained by
performing a T-duality transformation based on one of the original
solution's symmetries~\cite{buscher}.  As applied to the translation
symmetry, $r \to r + \hbox{constant}$, of the Kerr S-brane solution,
the new solution which is generated in this way is given by the fields
$\tilde{g}_{\mu\nu}$, $\tilde{B}_{\mu\nu}$ and $\tilde{\Phi}$, which
are related to the Kerr S-brane solution (in the string frame) by the
explicit transformation
\begin{eqnarray}\label{duality4dim}
\tilde g_{r r} &=& \frac{1}{g_{r r}}\,, 
\qquad
\tilde g_{r \alpha} = \frac{B_{r \alpha}}{g_{r r}}\,, 
\qquad 
\tilde g_{\alpha\beta} = g_{\alpha\beta} - 
\frac{(g_{r \alpha} g_{r \beta} - B_{r\alpha} B_{r\beta})}{g_{rr}}\,, 
\nonumber \\
\tilde B_{r\alpha} &=& \frac{g_{r\alpha}}{g_{rr}}\,, 
\qquad 
\tilde B_{\alpha\beta}=B_{\alpha\beta}
-\frac{2g_{r[\alpha} B_{\beta]r}}{g_{rr}}\,,
\nonumber\\ 
\tilde \Phi &=&  \Phi - \ln g_{rr} \,, 
\end{eqnarray}
where $\alpha$ and $\beta$ denote all of the coordinate directions
except for $r$.

This transformation furnishes us with the following new solution for
the action~(\ref{het4dim}), describing a time dependent configuration
in the presence of nontrivial dilaton and antisymmetric tensor field:
\begin{eqnarray}\label{solaftdual4d}
    d \tilde{s}^{2} &=&
    -\frac{\Sigma}{\Delta} d t^{2} + \frac{\Sigma}{\Delta+a^{2}
    \sinh^{2}\theta} d r^{2} +\Sigma \, d \theta^{2} + \frac{\Delta
    \Sigma \sinh^{2}\theta}{\Delta
    +a^{2} \sinh^{2}\theta} d \phi^{2}\,, 
\nonumber \\
    \tilde \Phi &=&  -\ln \left[ \frac{\Delta+a^{2}
    \sinh^{2}\theta}{\Sigma} \right],  
\nonumber\\ 
    \tilde{B}_{\phi r} &=& - \frac{2 m a t \sinh^{2}\theta}{\Delta
    +a^{2} \sinh^{2}\theta}\,.
\end{eqnarray}

\FIGURE{{\epsfig{file=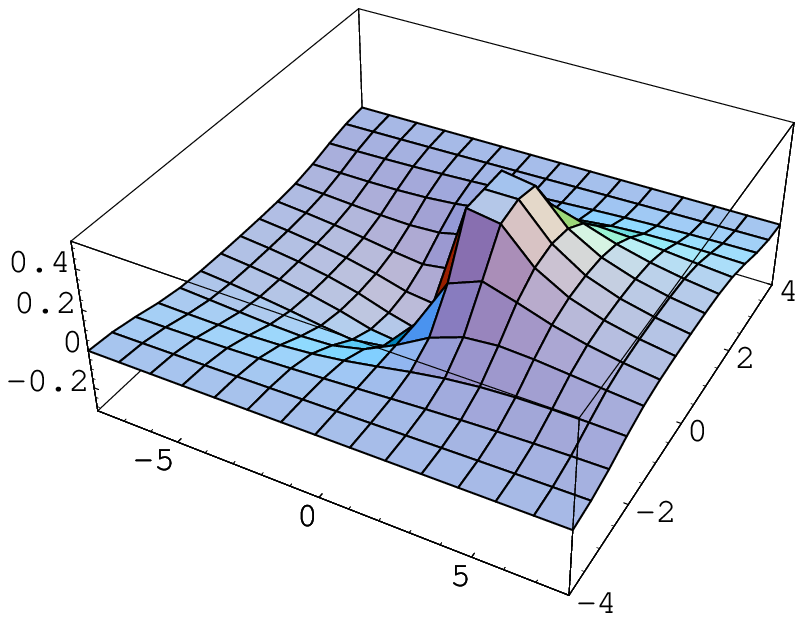, width=.45\textwidth}}%
\caption{Behavior of the scalar field for the solution in
  eq.~(\ref{solaftdual4d}), for $m=1$, $a=1.2$.\label{scalpict3D}}}

This configuration is completely time-de\-pen\-dent, and regular, provided
that $\Delta>0$, that is $a^{2} > m^{2}$, just as was the original
solution before performing the T-duality.  In this case, the parameter
$a$, whose size rules the regularity of the solution, is associated
with the ``charge'' of the non vanishing antisymmetric two form.  The
scalar field is also completely regular for this range of $a$, and it
depends on both the coordinates $t$ and $\theta$. The dilaton
configuration has a global maximum and a global minimum, and it
asymptotically vanishes: it is represented in figure~\ref{scalpict3D}.

For $a^2 < m^2$, the metric has Cauchy horizons at the surfaces
$\Delta=0$. The geometry is characterized by the presence of internal
stationary regions, containing CTC's, and curvature singularities when
the dilaton diverges, on the curves given by the solution of the
equation
\begin{equation}
t^{2}-2 m t +a^{2} \cosh^{2}{\theta}=0\,.
\end{equation}

For this four dimensional solution with $a^2 < m^2$, in the presence
of non trivial dilaton, it is not possible to find a configuration
free of curvature singularities. We will show, in
subsection~\ref{nscstring}, how this condition can be reached, in a
higher dimensional example, switching on additional components of the
antisymmetric tensor field.

\subsection{Higher dimensional S-Kerr solutions}

The vacuum solution in four dimensions that we discussed in the first
part of this letter, can be generalized to $d$ dimensional vacuum
solutions, at least in two different ways. The first, immediate way,
is to simply add to the solution ($d-4$) flat space-like directions
(we will use this observation in the next subsection as starting point
to find non trivial configurations via solution generating
techniques).  More interestingly, we can add ($d-4$) extra dimensions,
characterized by a hyperbolic symmetry.  This can be done by applying
a suitable analytic continuation to the higher dimensional Kerr black
hole with only one rotation parameter, which was found in
ref.~\cite{myersperry}.  The stationary solution, starting point for
our discussion, is given by
\begin{eqnarray}\label{perry}
ds_{d}^2 &=& -\frac{\Delta - a^2\,\sin^2\theta}{\Sigma}\,dt^2 + 
   \frac{\Sigma}{\Delta}\,dr^2 + \Sigma\, d\theta^2 
 + \frac{4\,m\,a\,r^{5-d}\,\sin^2\theta}{\Sigma}\,dt\,d\phi  + 
\nonumber\\
&& + \frac{\sin^2\theta}{\Sigma }[ (r^2 +a^2)^2 - 
             \Delta\,a^2\,\sin^2\theta]  d\phi^2 
    + r^2\cos^2\theta\, d\Omega_{d-4}^2\,,
\end{eqnarray}
where 
\begin{eqnarray}
\Delta &=& r^2 + a^2 - 2\,m\, r^{5-d} 
\nonumber \\
\Sigma &=& r^2 + a^2\,\cos^2\theta 
\end{eqnarray}
and $d\Omega_{d-4}$ is the line element of the ($d-4$) sphere. The
horizons are located at the points where $\Delta =0$.

Consider now the following complex replacement
\begin{equation} 
t \to i r\,, 
\qquad 
r \to it\,, 
\qquad 
\theta \to i\theta\,, 
\qquad 
a \to ia\,, 
\qquad 
m \to i^{d-3} m
\qquad 
d\Omega_{d-4} \to i d{\mathcal H}_{d-4} 
\end{equation}
where $d\Omega_p$ and $d{\mathcal H}_p$ respectively denote the line
elements for the $p$-sphere and $p$-hyperbola. Applied to the solution
eq.~(\ref{perry}) this transformation leads to the following
time-dependent solution in $d$ dimensions:
\begin{eqnarray}\label{timerotatingdd}
ds_{d}^2 &=& -\frac{\Sigma}{\Delta}\,dt^2 + 
   \frac{\Delta + a^2\,\sinh^2\theta}{\Sigma}\,dr^2 + \Sigma\, d\theta^2 
 - \frac{4\,m\,a\,t^{5-d}\,\sinh^2\theta}{\Sigma}\,dr\,d\phi  + 
\nonumber\\
&& + \frac{\sinh^2\theta}{\Sigma }[ (t^2 +a^2)^2 +
             \Delta\,a^2\,\sinh^2\theta] \, d\phi^2 
    + t^2\cosh^2\theta\, d{\mathcal H}_{d-4}^2\,,
\end{eqnarray}
where now
\begin{eqnarray}
\Delta &=& t^2 + a^2 - 2\,m\, t^{5-d} 
\nonumber \\
\Sigma &=& t^2 + a^2\,\cosh^2\theta \,.
\end{eqnarray}
This configuration has horizons at the locus of points defined by the
solutions to the equation
\begin{equation}\label{hordd}
    \Delta = t^2 + a^2 - 2m t^{5-d} = 0 
\qquad {\rm or} \qquad
    t^{d-3} + a^2 \,t^{d-5}- 2m =0  \,.
\end{equation}

From this equation is easy to see when, and how many, horizons are
present.  One finds that, in five dimensions, the geometry is
completely time dependent, without horizons or curvature
singularities, for $a^2>2m$. When this last inequality is not
satisfied, an internal, regular stationary region is present, which
contains CTC's.

Starting from six dimensions, the global structure changes
dramatically. Let us consider only the region for positive $t$, since
there is a curvature singularity at the origin.  Although the metric
is asymptotically time dependent, for positive $m$ it is characterized
by the presence of a Cauchy horizon, that separates the time-dependent
region from an internal, stationary region containing a naked
singularity at the origin. For negative $m$, the time dependent
geometry is characterized by an initial, space-like singularity, since
there are no horizons present.

Consequently, for six and more dimensions, our analytic continuation
applied to~(\ref{perry}) furnishes a singular geometry with a global
structure similar to previously known S-branes.

\subsection{A nonsingular charged S-string}\label{nscstring}

In this section we next construct a five dimensional configuration
which represents a solution to the field equations for the following
string-frame action in five dimensions:
\begin{equation}\label{het5}
S= -\int dx^5 \sqrt{-g_5} \,e^{-\Phi}\left[ R + (\nabla\Phi)^2
          - \frac{1}{12}\,H^2_\emph{3}\right].
\end{equation}
As before, $\Phi$ is the dilaton while $H_\emph{3}$ the antisymmetric
3-form field strength: $H_\emph{3} = \hbox{d} B_{\it 2}$.

It is relatively easy to find new solutions for this system. In order
to find them we proceed in an analogous way to that outlined
in~\cite{hhs}, using the following three steps.

\paragraph{Step 1:} we start with the vacuum spacetime of
section (2.2) --- i.e.\ eq.~(\ref{kerrSbr}) --- and we add to it an
extra flat direction,\footnote{We could easily add $n$ flat
  directions, but we concentrate on one for the sake of clarity.}
whose coordinate we call $x$. The solution obtained in this way
corresponds to a time-dependent S-string configuration, and is given
by
\begin{eqnarray}\label{stringkerrSbr}
    d s^{2} &=&  -\frac{\Sigma}{\Delta} dt^{2}+ \left( \frac{\Delta+a^{2}
    \sinh^{2}\theta}{\Sigma}
    \right)d r^{2}+\Sigma d\theta^{2}+ 
\nonumber \\
&&+  \left[(t^2+a^{2})^{2}+a^{2} \sinh^{2}\theta \Delta  \right]
    \frac{\sinh^{2}\theta}{\Sigma} d \phi^{2}
    -\frac{4 m\, a\, t}{\Sigma}\sinh^{2}\theta d \phi d r +d x^2
\nonumber \\
    \Phi &=& 0 
\nonumber \\
    B & = & 0 \,,
\end{eqnarray}
where we use the same conventions as in
subsection~\ref{simplerot}. This solution has the same global
properties as that of the 4-dimensional SKerr solution.

\paragraph{Step 2:} we  rotate the solution in the $(r,x)$
plane, by the transformation
\begin{equation}\label{rottran}
    r = \hat r \cos\alpha +\hat x \sin \alpha \,, 
\qquad 
    x =  \hat x \cos\alpha - \hat r \sin \alpha \,.
\end{equation}
We obtain the following rotated solution
\begin{eqnarray}\label{rottimedep}
    d\hat s_5^2 & = & - \frac{\Sigma}{\Delta}\,dt^2
    + \left[ \frac{\Delta + a^2\sinh^2 \theta + 2mt\sin^2\alpha}
       {\Sigma} \right] d\hat r^2
    + \left[\frac{\Delta + a^2 \sinh^2\theta
    + 2mt\cos^2\alpha}{\Sigma}\right] d\hat x^2  +
\nonumber \\
&& + \Sigma\, d\theta^2  + \frac{\sinh^2\theta}{\Sigma} \left[ (t^2+ a^2)^2
     + a^2\,\sinh^2\theta \Delta \right]\, d\phi^2
    - \frac{4\,m\,t\,\cos\alpha\sin\alpha}{\Sigma}\, d\hat x\, d\hat r -
\nonumber\\
&& -\frac{4\,m\,a\,t\,\sinh^2\theta\cos\alpha}{\Sigma}\, d\phi\, d\hat r
- \frac{4\,m\,a\,t\,\sinh^2\theta\sin\alpha}{\Sigma}\, d\phi\, d\hat x
\nonumber \\
\Phi &=& 0 
\nonumber \\
    B & = & 0 \,.
\end{eqnarray}

\paragraph{Step 3:} we  apply the  T-duality
transformation eq.~(\ref{duality4dim}) to the previous
solution~(\ref{rottimedep}), using the symmetry associated with
translations along the direction $\hat x$. We obtain the following new
metric
\begin{eqnarray}\label{cuerda1}
    d\tilde s_5^2 & = & -  \frac{\Sigma}{\Delta}\,dt^2 + \Sigma \,d \theta^2 +
    \frac{\Delta + a^2 \sinh^2\theta}{\Delta
                    + a^2\,\sinh^2\theta + 2\,m\,t\,\cos^2\alpha}\, d\hat r^2
    + \tilde g_{\phi\phi}\, d\phi^2  +
\nonumber \\  
&& + \frac{\Sigma}{\Delta + a^2 \sinh^2\theta + 2\,m\,t\,\cos^2\alpha}
\,d\hat x^2 - \frac{4\,m\,a\,t\,\sinh^2\theta\, \cos\alpha}{
\Delta + a^2 \,\sinh^2\theta +2\,m\,t\cos^2\alpha}\,d\hat r\, d\phi\,, 
\qquad
\end{eqnarray}
where $\tilde g_{\phi\phi}$ is given by the expression
\begin{equation}
    \tilde g_{\phi\phi} =
 \frac{\sinh^2(\theta)}{\Delta + a^2 \,\sinh^2\theta 
+ 2\,m\,t\,\cos^2\alpha}
    \left[ \Delta\,\Sigma +  2\,m\,t\,\cos^2\alpha\,(t^2 + a^2) \right].
\end{equation}
The dilaton and antisymmetric form are similarly given by
\begin{eqnarray}
    \tilde B_{\hat x\hat r} &=& -\frac{2\,m\,t\,\cos\alpha\, \sin\alpha}
       {\Delta + a^2 \sinh^2\theta + 2\,m\,t\, \cos^2\alpha}\,,   
\\
\tilde B_{\hat x\hat \phi} &=& -\frac{2\,m\,a\,t\,\sin\alpha\,\sinh^2\theta}
{\Delta + a^2 \sinh^2\theta + 2\,m\,t\, \cos^2\alpha }\,, 
\\
\tilde \Phi &=& - \ln{\frac{[\Delta + a^2 \sinh^2\theta 
+ 2\,m\,t\, \cos^2\alpha]}{\Sigma}}\,.
\end{eqnarray}

Taking the limit $\alpha=0$, one recover the
solution~(\ref{stringkerrSbr}).  In the limit $\alpha=\frac{\pi}{2}$,
instead, one finds essentially the T-dual solution discussed in
section~\ref{Tdual4}, with the addition of an extra flat dimension.

The charged string configuration coupled to the antisymmetric tensor
field constructed above is completely time dependent and regular
for $a^2 > m^2$ (that is, $\Delta$ is always positive).  When $a^2 <
m^2$, internal, stationary regions are present, separated by Cauchy
horizons from the external time dependent regions. In this case,
curvature singularities are possibly present when the dilaton
diverges, inside the stationary region, at points satisfying the
equation
\begin{equation}\label{cpsingstr}
\Delta + a^2\,\sinh^2\theta + 2mt\,\cos^2\alpha = 0\,.
\end{equation}

For this configuration, the presence of additional non zero components
for the field $B$, in comparison to the simpler four dimensional case
discussed in section~\ref{Tdual4}, allows to obtain non singular
solutions with horizons and non trivial dilaton: it is enough to
choose the rotation angle $\alpha$ in such a way that the following
inequalities are satisfied
\begin{equation}
0 < \sin^{4}{\alpha}< \frac{a^2}{m^{2}}\,,
\end{equation} 
as an analysis of the equation~(\ref{cpsingstr}) can easily show.  The
internal, regular stationary regions, however, contain again CTC's.

\subsection{SKerr in heterotic string theory}

In this final section, we generate new solutions using the method of
twisting in heterotic string theory, as introduced by Hassan and
Sen~\cite{hassansen}. In that work, a class of symmetries of the
solutions to heterotic string theory was discovered for configurations
which are independent of $d$ of the space-time directions, and for
which the background gauge fields are neutral under $p$ of the $\U(1)$
generators of the gauge group. The symmetry identified by these
authors is $O(d)\times O(d+p)$ if all of the $d$ directions are
space-like, or $O(d-1,1)\times O(d+p-1,1)$, if the signature of the
$d$-coordinates is lorentzian.

For our purposes, our interest in this symmetry is to generate new
inequivalent time dependent solutions from known ones. Indeed,
in~\cite{hassansen}, the authors applied this transformation to a
static 6-brane in ten dimensions, to generate a new solution carrying
electric and magnetic charges, antisymmetric tensor field charge and
non trivial dilaton field. In~\cite{senhete}, Sen derived in a similar
way a rotating, charged black hole with nontrivial dilaton, magnetic
fields and antisymmetric tensor field in heterotic string theory,
starting from a rotating solution in the vacuum.

We are now interested in applying these transformations to the
5-dimensional Kerr S-string constructed adding a flat direction, $x$,
to the 4-dimensional Kerr S-brane solution (our starting point is the
first step in the previous subsection, that is, the solution in
eq.~(\ref{stringkerrSbr})); our discussion will be based
on~\cite{maha}.  Thus we start with the low-energy action for
heterotic string theory in $D=5$ dimensions, with the other $(10-D)$
dimensions compactified. We do not include the massless \pagebreak[3]
compactification moduli in the effective action as well as higher
derivative terms. Then, we are left with the 5-dimensional metric, the
dilaton, the antisymmetric two form field and the Maxwell field, whose
action is given~by
\begin{equation}\label{hete5}
    S= -\int dx^5 \sqrt{-g_5} \,e^{-\Phi}\left[\, R + (\nabla\Phi)^2
        - \frac{1}{12}\,H^2_\emph{3} -\frac{1}{8}\, F^{2}_\emph{2}\right] .
\end{equation} 
Here $\Phi$ is the dilaton field, $F_{\mu\nu} = \partial_\mu\,A_\nu
-\partial_\nu\, A_\mu$ is the field strength of the $\U(1)$ gauge
field. The three form field $H$ is given by $H_{\mu\nu\rho} =
\partial_\mu\,B_{\nu\rho}$ + cyclic permutations - $CS$, where the
$\U(1)$ Chern-Simons term is $CS= \frac{1}{4}(A_\mu\, F_{\nu\rho} $ +
cyclic permutations).

We wish to apply the Hassan-Sen transformation to the solution in
eq.~(\ref{stringkerrSbr}), which is independent of the two spatial
coordinates, $(r,x)$ (that is, $d=2$). Given a solution
($g_{\mu\nu},\,\Phi,\,A_{\mu},\,B_{\mu\nu}$), a transformed solution
($\tilde g_{\mu\nu},\,\tilde\Phi,\,\tilde A_{\mu},\,\tilde
B_{\mu\nu}$) can be obtained in the following way.

Define the $11\times 11$ matrix $M$ as
\begin{eqnarray} \label{M}
  M = \left(
      \begin{array}{ccccc}
        (K^T - \eta){g}^{-1}(K - \eta)
        && (K^T - \eta){g}^{-1}(K + \eta)
    && -(K^T-\eta){g}^{-1}{A} \\[4pt]
    (K^T + \eta){g}^{-1}(K - \eta)
    && (K^T + \eta){g}^{-1}(K + \eta)
    && -(K^T+\eta){g}^{-1}{A} \\[4pt]
        -{A}^T\hat{g}^{-1}(K-\eta)
    && -{A}^T {g}^{-1}(K+\eta)
    && {A}^T {g}^{-1}{A}
      \end{array}
      \right) ,
\end{eqnarray}
where $K_{\mu\nu}$ is a $5\times 5$ matrix defined in terms of the
original solution,
\begin{equation} \label{K}
    K_{\mu\nu} = - {B}_{\mu\nu} - {g}_{\mu\nu}
           - \frac{1}{4}{A}_\mu {A}_\nu\,,
\end{equation}
and $\eta = $diag$(1,1,1,1,1)$.\footnote{In the case in which one of
  the coordinates in $d$ is time-like, $\eta=(-1,1,\dots,1)$.}

Our original solution~(\ref{stringkerrSbr}) has $A_\mu =0$, and so we
have $d=2$ and $p=1$. Thus the symmetry of the space of solutions is
$O(2)\times O(3)$. The Hassan-Sen transformation then acts on the
original solution, contained in the matrix $M$, to give a new
solution:
\begin{eqnarray}
    \tilde {M} & = & \Omega M \Omega^T 
\label{defNewM} \\
    \tilde\Phi & = & \Phi + \ln{\sqrt{\frac{\det{\tilde{g}}}{\det{g}}}} \,,
\end{eqnarray}
where $\Omega \in O(2)\times O(3)$ is given explicitly by
\begin{eqnarray}
  \Omega = \left(
           \begin{array}{cccc}
    I_3 &   &     &    \\
        & S &     &     \\
        &   & I_3 &   \\
            &   &     & R
           \end{array}
           \right) .
\end{eqnarray}
Here, $I_3$ represents the identity, while $R\in O(3)$ and $S\in
O(2)$. We can choose $R$ to be of the form
\begin{eqnarray} \label{R}
  R = \left(
      \begin{array}{ccc}
      \cos{\alpha_2} & \sin{\alpha_2} & 0 \\
     - \sin{\alpha_2} & \cos{\alpha_2} & 0 \\
      0            & 0            & 1
      \end{array}
      \right)
      \left(
      \begin{array}{ccc}
      \cos{\alpha_1} & 0 & \sin{\alpha_1} \\
      0            & 1 & 0            \\
      -\sin{\alpha_1} & 0 & \cos{\alpha_1}
      \end{array}
      \right) ,
\end{eqnarray}
where $\alpha_1$ parameterizes rotations which mix the $r$ direction
with the internal coordinate, and $\alpha_2$ parameterizes rotations
in the ($r,x$) space. The $O(2)$-transformations are rotations of the
solution in the $(r,x)$-plane and we choose $S$ to be the identity
matrix.

With this choice for $\Omega$, applied to the matrix $M$ constructed
in terms of the original solution, we can determine the new solution
completely. The new metric is given by
\begin{eqnarray}
    d\tilde s^{ 2}_{5} &=& -{\Sigma \over \Delta}\,dt^2 + \Sigma\,d\theta^2
                  + dx^2 
    +{[\Sigma(\Delta +a^2\sinh^2\theta) 
     +\sin^2\alpha_2\cos^2\alpha_1\,m^2\,t^2]
    \over [m\,t\,(1- \cos\alpha_1\cos\alpha_2) -\Sigma]^2} \,dr^2 +
\nonumber \\
    &&  + \tilde g_{\phi \phi} \, d\phi^2
    + {2\,m\, t \,a\sinh^2\theta [\sin^2\alpha_2 \cos^2\alpha_1 \,m\,t -
                   (1+ \cos\alpha_1 \cos\alpha_2 ) \Sigma]
    \over [\Sigma -m \,t\, (1-\cos\alpha_1 \cos\alpha_2)]^2}\, dr d\phi +
\nonumber \\
   &&+ {2\,m\,t\,\sin\alpha_2 \cos\alpha_1 \over
     [m \,t \,(1-\cos\alpha_1 \cos\alpha_2) - \Sigma]} \, dx dr +
    {2\,m\,t\,\sin\alpha_2 \cos\alpha_1\, \sinh^2\theta \over
        [m \,t \,(1-\cos\alpha_1 \cos\alpha_2) - \Sigma]} \, dx d \phi\,,\qquad
\end{eqnarray}
where
\begin{eqnarray}
    \tilde g_{\phi \phi} &=& \frac{\sinh^2\theta}{\Sigma} \Bigg\{(t^2+a^2)^2 +
             a^2\sinh^2\theta \Delta
    + m^2 t^2 a^2\sinh^2\theta\times
\nonumber \\
&&\hphantom{\frac{\sinh^2\theta}{\Sigma} \Bigg\{}\! 
\times\Bigg[\frac{2 (1-\cos\alpha_1 \cos\alpha_2)}{m \,t (1
 -\cos\alpha_1 \cos\alpha_2) - \Sigma}  -\frac{\sin^2\alpha_1 \Sigma}{[m \,t 
(1-\cos\alpha_1 \cos\alpha_2)-\Sigma]^2} \Bigg] \Bigg\} \,.\qquad
\end{eqnarray}

The dilaton and background fields are similarly given by
\begin{eqnarray}
\tilde \Phi &=& -\ln\left[\frac{\Sigma -
   m t (1-\cos\alpha_1 \cos\alpha_2)}{\Sigma}\right], 
\nonumber \\
\tilde A_\phi &=& -{2m t a\sin \alpha_1\sinh^2\theta\over \Sigma
    -m t (1- \cos\alpha_1 \cos\alpha_2)}  \,,    
\nonumber \\ 
\tilde A_r &=& -{2m t \sin\alpha_1 \over \Sigma
    -m t (1- \cos\alpha_1 \cos\alpha_2)}   \,,    
\nonumber \\ 
\tilde B_{r\phi} &=& {m t a (1- \cos\alpha_1 \cos\alpha_2)\sinh^2\theta \over
    \Sigma -m t (1- \cos\alpha_1 \cos\alpha_2)}  \,, 
\nonumber \\ 
\tilde B_{rx} &=& {m t \sin\alpha_2 \cos\alpha_1 \over
    \Sigma -m t (1- \cos\alpha_1 \cos\alpha_2)}  
\nonumber \\ 
\tilde B_{x \phi} &=& {- m t a \sin\alpha_2 \cos\alpha_1\sinh^2\theta \over
    \Sigma -m t (1- \cos\alpha_1 \cos\alpha_2)} \,,
\end{eqnarray}
while the other components of $\tilde A_\mu$ and $\tilde B_{\mu\nu}$
vanish.

This string-like, time dependent solution has a global structure which
resembles that of the charged S-string of last section. We have a
completely time-dependent, regular, causally well defined
configuration whenever $a^{2}>m^{2}$. In the opposite case, there
could be closed timelike curves in the internal region and naked
singularities situated where the dilaton, the $\tilde A$ and $\tilde
B$ fields are singular. This happens if the following equation has
real roots
\begin{equation}
    \Sigma - m \,t \,(1-\cos\alpha_1 \cos\alpha_2)=0\,.
\end{equation}
Also in this case, for $a^2<m^2$, one can judiciously choose the
angles $\alpha_1$ and $\alpha_2$, in such a way that the metric is
singularity free, although the internal stationary regions contain
CTC's.

Let us end noticing that the solution we obtained in this section
\emph{via} a Hassan-Sen transformation, is actually connected to the
previous solutions for certain choices of the angles $\alpha_i$.  For
$\alpha_1=\alpha_2=0$, one naturally finds the vacuum solution of
eq.~(\ref{stringkerrSbr}). More interestingly, consider the case
$\alpha_1=0$, and $\alpha_2=2 \alpha$.  One can explicitly check that
in this way the resulting solution reduces to the charged solution we
presented in the third step of section~\ref{nscstring}, once we
perform on the latter a new coordinate rotation~(\ref{rottran}), with
the same angle $\alpha$.

\section{Conclusions}

In the present letter we have identified several classes of exact
time-dependent solutions to the low-energy supergravity equations in
four and higher dimensions. Unlike many of the S-brane solutions
discussed earlier in the literature, some of the new solutions we
obtain are particularly interesting because they are nonsingular,
inasmuch as they have neither horizons, curvature singularities or
closed timelike curves.

\looseness=1 We discuss the simplest such solution --- the Kerr
S-brane --- in some detail. This is obtained from the Kerr solution by
making complex substitutions of coordinates and parameters in
precisely the same manner as the more symmetric S-brane solutions have
been constructed from the Schwarzschild spacetime. The solution
contains two parameters, $a$ and $m$, and reduces to the earlier
S-brane configurations as $a \to 0$. We show that these solutions are
nonsingular and causally well behaved if they satisfy $a^2 > m^2$.
When $0< a^2 < m^2$, although the solution is still free of curvature
singularities, it is characterized by internal stationary regions that
contain closed time-like curves, which make them problematic from the
point of view of causality. It might be interesting to study issues
such as stability and particle production in these simple
non-singular, time dependent backgrounds.

\looseness=1 We then generalize the simplest solution to include more
dimensions, and non trivial configurations for the typical fields of
low-energy supergravity, applying various solution generating
techniques. We concentrate on S-string backgrounds in five dimensions,
that contain a scalar field, and antisymmetric one- and two-form
fields.  Also in these richer situations, we are able to identify time
dependent solutions that are completely regular in the entire
space-time.  Simple extensions of these five dimensional solutions to
higher dimensions can be obtained by taking a product with additional
flat directions.  Other generalizations can be achieved by performing
different kinds of duality transformations.
 
It would be interesting to understand the connection between the
tachyon field dynamics in string theory and the regular time dependent
solutions we have presented here.  Also, it is worth to investigate
the possibility to construct realistic cosmological models based on
these geometries.

\paragraph{Note added:} while this letter was in the last
stages of preparation, we received the preprint in
reference~\cite{wang}, which contains substantial overlap with the
first part of the present work.

\acknowledgments

We would like to thank O. Obreg\'on for useful discussions.  C.B. is
funded by NSERC (Canada), FCAR (Qu\'ebec) and McGill
University. F.Q. is partially funded by PPARC and the Royal Society
Wolfson award. G.T. is supported by the European TMR Networks
HPRN-CT-2000-00131, HPRN-CT-2000-00148 and HPRN-CT-2000-00152.
I.Z. is supported by the United States Department of Energy under
grant DE-FG02-91-ER-40672.

\end{document}